%% file: main.tex
\documentclass[10pt,reprint,nocopyrightspace,numbers]{sigplanconf}

\usepackage{mathptmx}
\usepackage{microtype}
\usepackage{bec}
\usepackage{tikz}
\usetikzlibrary{shapes,backgrounds,arrows,calc,automata,positioning}
\usepackage[underline=false]{pgf-umlsd}
\usepackage{paralist}

\tikzstyle{hmmstate}=[shape=circle,draw=blue!50,fill=blue!20]
\tikzstyle{hmminitialabove}=[initial,initial text={$I_\mathcal{H}:1$}, initial where=above]
\tikzstyle{hmminitialleft}=[initial,initial text={$I_\mathcal{H}:1$}, initial where=left]
\tikzstyle{hmminitialright}=[initial,initial text={$I_\mathcal{H}:1$}, initial where=right]
\tikzstyle{pfsainitial}=[initial,initial text={$I(x_1):1$}, initial where=above]
\tikzstyle{pfsainitialabove}=[initial,initial text={$I_\mathcal{A}:1$}, initial where=above]
\tikzstyle{pfsainitialleft}=[initial,initial text={$I_\mathcal{A}:1$}, initial where=left]
\tikzstyle{pfsainitialright}=[initial,initial text={$I_\mathcal{A}:1$}, initial where=right]
\tikzstyle{pfsafinal}=[hmmstate,accepting]
\tikzstyle{observation}=[shape=rectangle,draw=orange!50,fill=orange!20]
\tikzstyle{lightedge}=[<-,dotted]
\tikzstyle{mainstate}=[state,thick]
\tikzstyle{mainedge}=[<-,thick]

\renewcommand{\topfraction}{0.95}

\renewcommand{\floatpagefraction}{0.9}

\setboolean{showjedi}{false}

\begin{document}
\input{macros}

\setlength{\pdfpageheight}{\paperheight}
\setlength{\pdfpagewidth}{\paperwidth}

\conferenceinfo{CONF 'yy}{Month d--d, 20yy, City, ST, Country}
\copyrightyear{20yy}
\copyrightdata{978-1-nnnn-nnnn-n/yy/mm}
\copyrightdoi{nnnnnnn.nnnnnnn}



\title{Abstracting Event-Driven Systems with Lifestate Rules}

\authorinfo{Shawn Meier
\and Aleksandar Chakarov
\and Maxwell Russek
\\ Sergio Mover
\and Bor-Yuh Evan Chang%
}
           {University of Colorado Boulder}
           {\{shawn.meier, aleksandar.chakarov, maxwell.russek, sergio.mover, evan.chang\}@colorado.edu}

\maketitle

\begin{abstract}
\input{abstract}
\end{abstract}




\section{Introduction}
\label{sec:introduction}

\JEDI{Motivation: Challenge in developing against event-driven frameworks.}
We consider the problem of specifying and mining the object protocols used by event-driven software frameworks.
Programming against event-driven frameworks is hard. 
In such frameworks, programmers develop client applications~(\emph{apps}) against the \emph{framework} by implementing \emph{callback} interfaces that enable the application to be notified when an \emph{event} managed by the framework occurs (e.g., a user-interface~(UI) button is pressed). The app may then delegate back to the framework through method calls to the application programming interface~(API) (e.g., to direct a change in the UI display).
To develop working apps, the application programmer must understand the complex object protocols implemented by the event-driven framework. For example, the framework may guarantee particular ordering constraints on callback invocations (known as \emph{lifecycle} constraints), and the application programmer may have to respect particular orderings of API calls (i.e., \emph{typestate} constraints
).

\JEDI{Why important: Such frameworks are pervasive. Why hard: No specs about these constraints. Hard to get because of size of framework and native code.}
Unfortunately, such protocol specifications are complex to describe and maintain---and almost always incomplete. Because lifecycle constraints are so central to implementing apps, they are typically discussed in the framework documentation,
but they are incomplete enough that developers spend considerable manual effort to derive more complete specifications~\cite{xxv-androidlifecycle}.



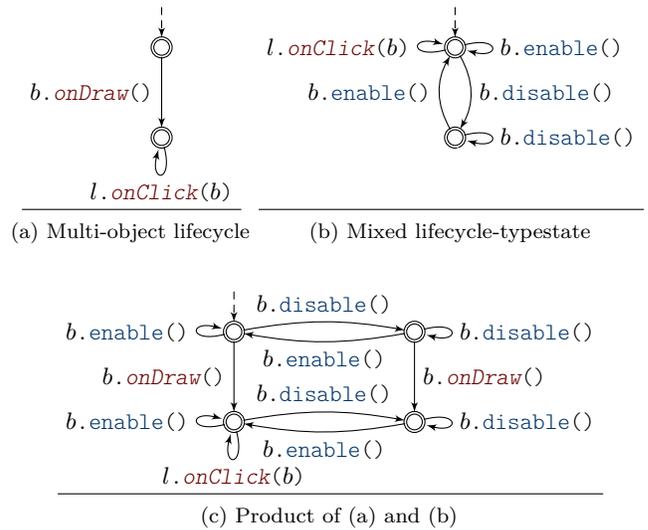
\begin{figure}\centering\small
\setlength{\fboxsep}{0pt}
\subfloat[Multi-object lifecycle]{\label{fig:intro-automata:lifecycle}
\begin{tikzpicture}[inner frame sep=0pt, show background bottom, node distance=4cm,auto,>=latex']
  \tikzstyle{every initial by arrow}=[densely dashed]
  \node [lastate, lainitial] (init) {};
  \node [lastate] (drawn) [below of=init] {};
  \path [->] (init) edge node [lalabel, left] {\codej{$b$.|cb:onDraw|()}} (drawn);
  \path [->] (drawn) edge [loop below] node [lalabel] {\codej{$l$.|cb:onClick|($b$)}} (drawn);
\end{tikzpicture}
}
\subfloat[Mixed lifecycle-typestate]{\label{fig:intro-automata:mixed}
\begin{tikzpicture}[inner frame sep=0pt, show background bottom,node distance=4cm,auto,>=latex']
  \tikzstyle{every initial by arrow}=[densely dashed]
  \node [lastate, lainitial] (init) {};
  \node [lastate] (disabled) [below of=init] {};
  \path [->] (init) edge [loop left] node [lalabel] {\codej{$l$.|cb:onClick|($b$)}} (init);
  \path [->] (init) edge [bend left] node [lalabel] {\codej{$b$.|ci:disable|()}} (disabled);
  \path [->] (disabled) edge [bend left] node [lalabel] {\codej{$b$.|ci:enable|()}} (init);
  \path [->] (init) edge [loop right] node [lalabel] {\codej{$b$.|ci:enable|()}} (init);
  \path [->] (disabled) edge [loop right] node [lalabel] {\codej{$b$.|ci:disable|()}} (disabled);

  \path [->] (disabled) edge [loop below, color=white] node [lalabel] {\phantom{\codej{$l$.|cb:onClick|($b$)}}} (disabled);
\end{tikzpicture}
}
\hfill
\subfloat[Product of (a) and (b)]{\label{fig:intro-automata:product}
\begin{tikzpicture}[inner frame sep=0pt, show background bottom,node distance=4cm,auto,>=latex']
  \tikzstyle{every initial by arrow}=[densely dashed]
  \node [lastate, lainitial] (init) {};
  \node [lastate] (drawn) [below of=init] {};
  \node [lastate] (disabled) [right of=init, node distance=8cm] {};
  \node [lastate] (disableddrawn) [below of=disabled] {};

  \path [->] (init) edge node [lalabel, left] {\codej{$b$.|cb:onDraw|()}} (drawn);
  \path [->] (drawn) edge [loop below] node [lalabel] {\codej{$l$.|cb:onClick|($b$)}} (drawn);

  \path [->] (disabled) edge node [lalabel, right] {\codej{$b$.|cb:onDraw|()}} (disableddrawn);

  \path [->] (init) edge [bend left=10] node [lalabel] {\codej{$b$.|ci:disable|()}} (disabled);
  \path [->] (disabled) edge [bend left=10] node [lalabel] {\codej{$b$.|ci:enable|()}} (init);

  \path [->] (drawn) edge [bend left=10] node [lalabel] {\codej{$b$.|ci:disable|()}} (disableddrawn);
  \path [->] (disableddrawn) edge [bend left=10] node [lalabel] {\codej{$b$.|ci:enable|()}} (drawn);

  \path [->] (init) edge [loop left] node [lalabel] {\codej{$b$.|ci:enable|()}} (init);
  \path [->] (disabled) edge [loop right] node [lalabel] {\codej{$b$.|ci:disable|()}} (disabled);

  \path [->] (drawn) edge [loop left] node [lalabel] {\codej{$b$.|ci:enable|()}} (drawn);
  \path [->] (disableddrawn) edge [loop right] node [lalabel] {\codej{$b$.|ci:disable|()}} (disableddrawn);
\end{tikzpicture}
}
\caption{Event-driven object protocols are complex.}
\label{fig:intro-automata}
\end{figure}

\JEDI{Why hard: The automata explodes, exponential in the number of ``actions" or ``operations" or ``messages".}
To get a sense for the complexity of event-driven object protocols, consider the automaton-based specifications shown in \figref{intro-automata}.
The meaning of such a specification is that execution traces (projected on to the actions of interest) must be words accepted by the automaton.
In \figref{intro-automata:lifecycle}, we describe a portion of a lifecycle specification for a protocol between a button object $b$ and its click-listener $l$.
This automaton states that a \codej{$b$.|cb:onDraw|()} callback notification happens before any \codej{$l$.|cb:onClick|($b$)} callback invocations.
This specification is a multi-object lifecycle constraint because it orders callbacks on objects $b$ and $l$.

In \figref{intro-automata:mixed}, we show another important property of buttons and their click-listeners: a button can be enabled or disabled via API calls \codej{$b$.|ci:enable|()} and \codej{$b$.|ci:disable|()}, respectively.
When button $b$ is enabled (the upper state), the click-listener $l$ may receive the \codej{$l$.|cb:onClick|($b$)} callback notification. Through an API call to \codej{$b$.|ci:disable|()}, button $b$ becomes disabled. When button $b$ is disabled (the lower state), the click-listener $l$ can no longer receive \codej{$l$.|cb:onClick|($b$)} notifications.
This specification mixes a lifecycle (or callback-ordering) constraint with a typestate (or API call-ordering) constraint.

We call API calls \underline{in}to the framework, \emph{callins}, to clearly contrast them with \emph{callbacks} that are calls \underline{back} from the framework.
The specification in \figref{intro-automata:mixed} mixes two kinds of invocations: (1) events labeled by the callbacks they invoke to notify the app (written {\lstbasicstyle\cbstyle slanted}) and (2) callins that the app invokes to change the state of the framework (written {\lstbasicstyle\cistyle upright}).

\JEDI{Automata specifications are unwieldy, not compositional, and too precise.}
From just the examples in \figref{intro-automata}, we see that the specification space is rich and complex with multiple interacting objects over callbacks and callins.
Worse, the composition of the \codej{|cb:onDraw|}-before-\codej{|cb:onClick|} specification from \subref{fig:intro-automata:lifecycle} and the \codej{|ci:enable|}-\codej{|ci:disable|}-\codej{|cb:onClick|} specification from \subref{fig:intro-automata:mixed} is the product automaton shown in \subref{fig:intro-automata:product}.
In this product automaton, the left states are where button $b$ is enabled versus the right states where $b$ is disabled, and the upper states are where button $b$ has yet to be drawn versus the lower states are where $b$ has been drawn.
While automata are descriptive and natural, the \emph{key observation of this paper} is that automata are more precise than necessary or desired.

\JEDI{Contribution: a new kind of specification that is compositional.}
Thus, we take a departure from traditional automaton specifications and define a more abstract specification language consisting of \emph{lifestate rules} that captures the critical lifecycle and typestate constraints without representing them as automata.
We shall see that this specification language is more compact than automata and exposes lifecycle and typestate constraints as duals.
At a high level, lifestate rules focus on changes in the internal, hidden state of the event-driven framework.
For example, the  \codej{|ci:enable|}-\codej{|ci:disable|}-\codej{|cb:onClick|} specification from \figref{intro-automata:mixed} can be summarized as the following rules:
\[\begin{array}{lll}
\codej{$b$.|ci:enable|()}  & \text{enables}  & \codej{$l$.|cb:onClick|($b$)} \;\text{for some $l$}
\\
\codej{$b$.|ci:disable|()} & \text{disables} & \codej{$l$.|cb:onClick|($b$)} \;\text{for some $l$}
\end{array}\]
The first rule says that whenever the invocation \codej{$b$.|ci:enable|()} fires, the state of the framework changes to enable the event that invokes the \codej{$l$.|cb:onClick|($b$)} callback, and the second rule says when \codej{$b$.|ci:disable|()} fires, the state changes to disable the event that invokes \codej{$l$.|cb:onClick|($b$)}.
These rules do not enumerate the explicit invocation orders as in the automaton from \figref{intro-automata:mixed}.

Having defined lifestate rules for event-driven object protocols, we investigate specification mining techniques from dynamic traces of apps using a framework. To address the technical challenges to this end, we make the following contributions:
\begin{itemize}\itemsep 0pt
  \item We formalize a concrete semantics for event-driven systems using an abstract machine model \semname that captures an appropriate model of the internal, hidden state of the framework (\secref{concrete}). We introduce the notion ``allowedness'' for API callins, which we shall see is parallel to ``enabledness'' for events.
  \item We define a language for \emph{lifestate rules}, which incorporates mixed lifecycle-typestate constraints (\secref{specification}).
  \item We introduce \emph{callback-driven trace slicing}, a form of trace slicing~\cite{DBLP:conf/tacas/ChenR09} adapted to events and callbacks
(\secref{slicing}). 
The \semname{} model and this trace slicing algorithm is the basis for our dynamic analysis tool, \toolname, that records events, callbacks, and callins in running Android applications.
  \item We apply specification mining techniques to derive lifestate specifications (\secref{mining}). In particular, we apply unsupervised automata-based learning techniques based on probabilistic finite state automata
and hidden Markov models.
We also develop a direct lifestate mining technique based on propositional model counting (\sharpsat). The most significant challenge with lifestate mining is that the \emph{enables}, \emph{disables}, \emph{allows}, and \emph{disallows} that explain callback and callin invocations are not observable in concrete executions.
\item We empirically evaluate our mining algorithms on 133 traces from a corpus of 1,909 Android apps (\secref{evaluation}). Our results show that we can in fact learn several rules corresponding to actual Android behavior and one of which that was not in the Android documentation.
\end{itemize}

\section{Overview}
\label{sec:overview}

In this section, we motivate the challenges in specifying and mining
event-driven object protocols and then exhibit how our approach models
event-driven systems to mine lifestate rules about the Android framework.

We will show with our running example in \figref{feedremover-example} that reasoning about the correctness of apps requires a comprehensive understanding of the lifestate rules of various Android framework components.
In particular, the call marked with \danger{} can throw an \codej{IllegalStateException}.
This code is correct, but understanding why requires a specification of subtle relationships between the various callins and the events that trigger the callbacks.
We will also see that a slight change of this code makes it buggy.

\paragraph{Lifestate Specification.}
%
Android is a prominent example of an event-driven framework that dispatches a multitude of events (e.g., from user interaction or sensors).
An app gets notified about events by implementing callback interfaces that may then delegate back to the framework through the framework's callin interface. We consider an event-driven object protocol to be the rules governing the use of an interface consisting of callbacks and callins.
Unfortunately, the protocol rules are determined by the complex, internal behavior of the framework, which becomes a major source of confusion for app developers and thus of defects in apps.


\newsavebox{\SBoxSetEnabledTrue}
\begin{lrbox}{\SBoxSetEnabledTrue}\small
\codej{b.setEnabled(true)}
\end{lrbox}
\newsavebox{\SBoxSetEnabledFalse}
\begin{lrbox}{\SBoxSetEnabledFalse}\small
\codej{b.setEnabled(false)}
\end{lrbox}
\newsavebox{\SBoxFeedRemoverExample}
\begin{lrbox}{\SBoxFeedRemoverExample}\small
\lststopn\begin{lstlisting}[language=Java,alsolanguage=exthighlighting,style=number]
class FeedRemover extends AsyncTask {
 MainActivity a; Button b;
 void doInBackground() { $\ldots\;\text{\sl remove feed}\;\ldots$ }
 void |cb:onPostExecute|() {#\lstbeginn#
  a.remover = new FeedRemover(a, b);#\label{line-feedremover-alloc}#
  #\dashuline{\usebox{\SBoxSetEnabledTrue}}#;#\label{line-feedremover-enable}\lststopn#
 }
}
class MainActivity extends Activity {
 FeedRemover remover;
 void onCreate() {#\lststartn#
  Button b = $\ldots$;
  remover = new FeedRemover(this, b);#\label{line-feedremover-firstalloc}#
  b.setOnClickListener(new OnClickListener() {#\label{line-feedremover-listener}\lststopn#
   void onClick(View v) {#\lststartn#
    #\dashuline{\usebox{\SBoxSetEnabledFalse}}#; #\label{line-feedremover-disable}#
    remover.execute();   #\danger\label{line-feedremover-danger}\lststopn#
   }
  });
 }
}#\lststartn#
\end{lstlisting}
\end{lrbox}
\newsavebox{\SBoxAsyncTaskInterface}
\begin{lrbox}{\SBoxAsyncTaskInterface}\small
\begin{lstlisting}[language=Java,alsolanguage=exthighlighting]
abstract class AsyncTask {
 void |cb:onPostExecute|() { }
 abstract void doInBackground();
 final void execute() { $\ldots$ }
}
\end{lstlisting}
\end{lrbox}
\newsavebox{\SBoxAsyncExecute}\codejmk[\small]{\SBoxAsyncExecute}{(t:AsyncTask).execute()}
\newsavebox{\SBoxAsyncExecuteNoType}\codejmk[\small]{\SBoxAsyncExecuteNoType}{t.execute()}
\newsavebox{\SBoxAsyncPostExecute}\codejmk[\small]{\SBoxAsyncPostExecute}{t.|cb:onPostExecute|()}
\newsavebox{\SBoxButtonEnable}\codejmk[\small]{\SBoxButtonEnable}{(b:Button).setEnabled(true)}
\newsavebox{\SBoxButtonDisable}\codejmk[\small]{\SBoxButtonDisable}{(b:Button).setEnabled(false)}
\newsavebox{\SBoxListenerOnClick}\codejmk[\small]{\SBoxListenerOnClick}{(l:OnClickListener).onClick(b)}
\newsavebox{\SBoxAsyncInit}\codejmk[\small]{\SBoxAsyncInit}{(t:AsyncTask).<init>()}
%
\newsavebox{\SBoxFeedRemoverTrace}
\begin{lrbox}{\SBoxFeedRemoverTrace}\footnotesize
\begin{sequencediagram}
\newthread{f}{Framework\vphantom{App}}
\newinst[5]{a}{App\vphantom{Framework}}
\begin{scope}[>=triangle 60]
\begin{sdblock}{\fmtevt{Create}}{}
  \prelevel
  \begin{messcall}{f}{ \codej{(a:Activity).onCreate()} }{a}{}
    \begin{messcall}{a}{ \codej{(t1:AsyncTask).<init>()} }{f}{}
      \node[right=1mm] at (cf\thecallevel) {\lstnumberstyle\ref{line-feedremover-firstalloc}};
    \end{messcall}
    \prelevel
    \begin{messcall}{a}{ \codej{(b:Button).setOnClickListener(l:OnClickListener)} }{f}{}
      \node[right=1mm] at (cf\thecallevel) {\lstnumberstyle\ref{line-feedremover-listener}};
    \end{messcall}
    \prelevel
  \end{messcall}
  \prelevel
\end{sdblock}
\begin{sdblock}{\fmtevt{Click}}{}
  \prelevel
  \begin{messcall}{f}{ \codej{(l:OnClickListener).onClick(b:Button)} }{a}{}
    \begin{messcall}{a}{ \codej{(b:Button).setEnabled(false)} }{f}{}
      \node[right=1mm] at (cf\thecallevel) {\lstnumberstyle\ref{line-feedremover-disable}};
    \end{messcall}
    \prelevel
    \begin{messcall}{a}{ \codej{(t1:AsyncTask).execute()} }{f}{}
      \node[right=1mm] at (cf\thecallevel) {\lstnumberstyle\ref{line-feedremover-danger}};
    \end{messcall}
    \prelevel
  \end{messcall}
  \prelevel
\end{sdblock}
\begin{sdblock}{\fmtevt{PostExecute}}{}
  \prelevel
  \begin{messcall}{f}{ \codej{(t1:AsyncTask).|cb:onPostExecute|()} }{a}{}
    \begin{messcall}{a}{ \codej{(t2:AsyncTask).<init>()} }{f}{}
      \node[right=1mm] at (cf\thecallevel) {\lstnumberstyle\ref{line-feedremover-alloc}};
    \end{messcall}
    \prelevel
    \begin{messcall}{a}{ \codej{(b:Button).setEnabled(true)} }{f}{}
      \node[right=1mm] at (cf\thecallevel) {\lstnumberstyle\ref{line-feedremover-enable}};
    \end{messcall}
    \prelevel
  \end{messcall}
  \prelevel
\end{sdblock}
\end{scope}
\end{sequencediagram}
\end{lrbox} 

\begin{figure}\centering
\subfloat[The call \codej{remover.execute()} on \reftxt{line}{line-feedremover-danger} (marked with \danger) can throw an \codej{IllegalStateException} if the \codej{remover} task is already running.
This \codej{execute} call happens whenever the user clicks button \codej{b}.
This code is indeed safe because the developer enforces that only one instance of \codej{FeedRemover} is ever executing through the disabling and enabling of the button on lines~\ref{line-feedremover-disable} and~\ref{line-feedremover-enable}, respectively (shown with \protect\dashuline{dashed underline}).]{\label{fig:feedremover-code}\makebox[\linewidth]{\usebox{\SBoxFeedRemoverExample}}}
%
\par\subfloat[A \toolname{} trace recorded dynamically. Each box is the processing of an
event. Callbacks are depicted as arrows to the right from framework
to app code; callins are arrows to the left, which are labeled with
the corresponding program location from
\figref{feedremover-code}.]
{\label{fig:feedremover-trace}\makebox[\linewidth]{\scalebox{0.81}{\usebox{\SBoxFeedRemoverTrace}}}}
\caption{An example Android app whose safety depends on understanding the lifestate rules of two Android framework components \codej{AsyncTask} and \codej{Button}.
}
\label{fig:feedremover-example}
\end{figure}

To see this complexity, consider the code shown in \figref{feedremover-example}.
The app performs a time consuming operation (removing feeds) when the
user clicks the button \codej{b}. The operation is performed
asynchronously using \codej{AsyncTask} to not block the UI thread.
\figref{feedremover-trace} shows an execution trace of the app.
Execution time flows downwards. A box surrounds the callback and callin invocations that happen as the result of processing of a particular event.
First event \fmtevt{Create} invokes the callback \codej{(a:Activity).onCreate()} for the object \codej{a}. 
In this case, object \codej{a} has dynamic type \code{MainActivity}; for reasons we will discuss below, we label the object with its most precise supertype that is a framework type.
In the \codej{onCreate} callback, the code then invokes a callin for the \codej{AsyncTask} constructor (i.e., \codej{(t1:AsyncTask).<init>()}) and then the callin for \codej{setOnClickListener} following the app code at \reftxt{line}{line-feedremover-listener}.
Then, the trace shows the execution of other two events,
\fmtevt{Click}, with the callback \codej{|cb:onClick|}, and
\fmtevt{PostExecute}, with the callback \codej{|cb:onPostExecute|}.

\Figref{feedremover-trace} also depicts the back and forth flow of
control between framework and app code.
The boxes along the lines show the lifetime of activations.
In particular, we see that we can define a \emph{callback} invocation as the
invocation that transfers control from framework code to app code
(arrows to the right), while a \emph{callin} is the reverse (i.e., an
invocation that transfers control from app code to framework).
The difficulty in reasoning about the code in \figref{feedremover-code} is that framework behavior that governs the possible control flow shown in \figref{feedremover-trace} is obfuscated in the code.

Our running example is inspired by an actual
\href{https://github.com/AntennaPod/AntennaPod/issues/1304}{issue} in
AntennaPod~\cite{antennapodbug}
a podcast manager for Android with 100,000--500,000 installs from the
Google Play
Store
(a similar issue appears in the Facebook SDK for Android~\cite{facebookbug}).
The potential bug is that the callin \codej{AsyncTask.execute}, called
via \codej{remover.execute()} (\reftxt{line}{line-feedremover-danger}),
can throw an \codej{IllegalStateException}.

In the example the \codej{execute} call happens whenever the user
clicks button \codej{b}, which in turn triggers the event that dispatches to the
\codej{OnClickListener.onClick} callback on the click-listener (set on \reftxt{line}{line-feedremover-listener}). 
This code is indeed safe (i.e., does not throw the exception) because
the developer enforces that the execute method on \codej{FeedRemover}
is only called once. This property is enforced by the disabling and enabling of
button \codej{b} in
\codej{onClick}
and in \codej{|cb:onPostExecute|}, respectively.
Buttons in Android are disabled and enabled via calls to \codej{setEnabled(false)} and \codej{setEnabled(true)}, respectively (corresponding to \codej{disable} and \codej{enable} in \secref{introduction}).

To see why this button disabling and enabling enforces the single-executing-instance property, the developer needs to know the interactions between \codej{AsyncTask}s, \codej{Button}s, and \codej{OnClickListener}s in the Android framework.
Only with an understanding of the framework do we see that the call to \codej{b.setEnabled(false)} (\reftxt{line}{line-feedremover-disable}) disables the event that could trigger the \codej{onClick} callback (which calls the potentially problematic \codej{remover.execute()}).
Button \codej{b} (and thus the click event) can be enabled inside the callback \codej{|cb:onPostExecute|} (\reftxt{line}{line-feedremover-enable}), but again with an understanding of the framework, we see that \codej{|cb:onPostExecute|} is triggered by a post-execute event, which is in turn enabled by the call to \codej{remover.execute()} (\reftxt{line}{line-feedremover-danger}). When this happens, \codej{remover} is set to a newly-allocated instance of \codej{FeedRemover} (\reftxt{line}{line-feedremover-alloc}), which allows for the \codej{execute} callin to be executed safely. 
%

Without the calls to \codej{setEnabled} that disables and enables button \codej{b},
(on lines~\ref{line-feedremover-disable} and~\ref{line-feedremover-enable}, respectively),
the app is buggy because the user can click the button a second time, causing a second call to \codej{remover.execute()} on the same instance of \codej{FeedRemover}.

\begin{figure}
\begin{tabular}{@{}>{$}p{\linewidth}<{$}@{}}
\specDisallow{ \usebox{\SBoxAsyncExecute} \hfill}{ \;\usebox{\SBoxAsyncExecuteNoType} }
\\[1.5ex]
\specDisable{ \usebox{\SBoxButtonDisable} \\\hfill}{ \;\evtcb{Click}{\usebox{\SBoxListenerOnClick}} }
\\
\specEnable{ \usebox{\SBoxAsyncExecute} \\\hfill}{ \;\evtcb{PostExecute}{\usebox{\SBoxAsyncPostExecute}} }
\\
\specEnable{ \usebox{\SBoxButtonEnable} \\\hfill}{ \;\evtcb{Click}{\usebox{\SBoxListenerOnClick}} }
\\
\specAllow{ \usebox{\SBoxAsyncInit} \hfill}{ \;\usebox{\SBoxAsyncExecuteNoType} }
\end{tabular}
\caption{Some lifestate rules for \codej{AsyncTask} and \code{Button}.
The first rule specifies a typestate property, while the others are required to reason about the safety of the call to \codej{remover.execute()} in \figref{feedremover-code}.}
\label{fig:feedremover-rules}
\end{figure}

\newsavebox{\SBoxAsyncExecuteNormal}\codejmk{\SBoxAsyncExecuteNormal}{(t:AsyncTask).execute()}
\newsavebox{\SBoxAsyncExecuteNoTypeNormal}\codejmk{\SBoxAsyncExecuteNoTypeNormal}{t.execute()}

The ``understanding'' of the Android framework that is necessary to reason about the code in \figref{feedremover-code}  is precisely what we seek to capture with lifestate specifications.
In \figref{feedremover-rules}, we sketch the lifestate rules needed to reason about our running example. The first rule, 
\(\specDisallow{ \usebox{\SBoxAsyncExecuteNormal} }{ \;\usebox{\SBoxAsyncExecuteNoTypeNormal} }\)
says that when \codej{(t:AsyncTask).execute()} is invoked for an object \codej{t} of type \codej{AsyncTask}, the \codej{execute} callin on \codej{t} is \emph{{\bfseries dis}allowed}.
We use the terms \emph{allowed} and \emph{disallowed} to refer the state of callins; these terms are parallel to \emph{enabled} and \emph{disabled} for the state of events.
Overall, this rule specifies that it is erroneous to call \codej{execute} twice (without something else that re-allows it).

The subsequent three rules specify disabling (written $\nspecArrow\specDisableOp$) and enabling (written $\specArrow\specEnableOp$) events (e.g.  \codej{setEnabled} enables the \codej{onClick} callback).
To make clear that the enabled-disabled state refers to events, we give names to events, such as \fmtevt{Click} (written in \fmtevt{small caps}) and associate events with the callback(s) they trigger (written \evtcb{Event}{\codej{onCallback()}}).
Finally, the last rule states that the asynchronous task constructor \codej{(t:AsyncTask).<init>()} allows the \codej{t.execute()} callin.
We define and discuss the formal semantics of lifestate specifications in \secref{specification}.

\JEDI{We formalize the notion of enabled-disabled-allowed-disallowed messages in $\semname$.}
We argue that specifications describing the \emph{enabling-disabling of events} and the \emph{allowing-disallowing of callins} are central to event-driven object protocols.
The specifications capture the relations that informally say, ``When a particular method
invocation happens, the state of the framework changes to enable or
disable an event or to allow or disallow a callin.''
%
%
We unify the notions of events and callins by introducing the term \emph{message}.
To reason about lifestate specifications,
we formalize an abstraction of event-driven frameworks, called
$\semname$, that captures the internal state consisting of enabled and
allowed messages in \secref{concrete}.

\paragraph{Callback-Driven Trace Slicing and Lifestate Mining.}
%
%
%
We cannot actually observe the abstraction of the
internal framework state of a running Android app. Instead, we can
only observe the sequence of method invocations.
%
We use the operational semantics of $\semname$ to derive an
instrumented semantics that captures the behavior that we can observe:
messages classified into event, callback, callin, or internal
invocations. This instrumented semantics
specifies the recording needed to produce traces like the one shown in
\figref{feedremover-trace}
and is implemented in a tool called \toolname{}.

\newsavebox{\SBoxFeedRemoverProjected}
\begin{lrbox}{\SBoxFeedRemoverProjected}\small
\begin{lstlisting}[language=Java,alsolanguage=exthighlighting]
AsyncTask.<init>()
AsyncTask.execute()
|evt:PostExecute|$\frameop$AsyncTask.|cb:onPostExecute|()
\end{lstlisting}
\end{lrbox}
%

%
We are interested in rules about the interactions between events and callins on framework objects.
To collect traces across multiple executions, we apply the standard type abstraction to concrete objects, resulting in an abstract, signature message like \codej{AsyncTask.execute()}.

\JEDI{Don't want to mix different objects of the same type. Need to slice before abstraction.}
To obtain relevant lifestate rules and improve the feasibility of mining, we slice recorded traces into sub-traces that group together related method calls.
Our approach is roughly to slice a recorded trace into
sub-traces that collect method invocations whose arguments share a
particular concrete object, similar to
\citet{DBLP:conf/icse/PradelJAG12}.
However, we face the additional difficulty that an event message is
not directly a method invocation and the relevant framework object of
the event is typically hidden in the internal state of the message.
Our insight for identifying relevant event messages is that an event
should eventually trigger a callback where the relevant framework
object is also argument to the callback.
Thus, our \emph{callback-driven trace slicing} approach determines the relevance of an event message $\msg$ based on the arguments to the callback(s) that $\msg$ eventually triggers.
%
%
For example, in our trace slice for the concrete object \codej{t1:AsyncTask}, the slice includes the
\fmtevt{PostExecute} event because it eventually triggers the callback \codej{t1.|cb:onPostExecute|()} and so the
type-abstracted sub-trace for object \codej{t1} from the recorded trace in
\figref{feedremover-trace} is as follows:
\[
\usebox{\SBoxFeedRemoverProjected}
\]



Once we have sets of sliced traces, we apply and evaluate specification mining
techniques to learn lifestate rules.
As alluded to above, the
primary challenge for mining lifestate specifications is that
lifestate rules center around event enabling-disabling and callin
allowing-disallowing, which are not observable in recorded traces.
We present our mining approach in \secref{mining} and we evaluate
them in \secref{evaluation}.

\section{Modeling Event-Driven Programs}
\label{sec:concrete}

In this section, we formalize a small-step operational semantics for event-driven programs using an abstract machine model (\semname) that unifies \emph{enabledness} for event lifecycles and \emph{allowedness} for callin typestates.
%
Crucially, this semantics motivates lifestate specifications, as the concrete states of this semantics serves as the concrete domain from which lifestate specifications abstract (\secref{specification}).
This semantics also serves as a basis for describing our dynamic analysis for recording traces of events, callbacks, and callins (\secref{slicing}).

\begin{figure}[tb]\small
\begin{mathpar}
\begin{grammar}[][l@{}]
\expr \in \ExprSet
& \bnfdef & \val \bnfalt \cdots \bnfalt \enIf{\val}{\expr_1}{\expr_2} & primitives
\\
& \bnfalt & \enBind{\val_1}{\val_2} & thunks
\\
& \bnfalt & \enAllow{\val} \bnfalt \enDisallow{\val} \bnfalt \enInvoke{\val_1}{\val_2} & calls
\\
& \bnfalt & \enEnable{\val} \bnfalt \enDisable{\val} & events
\\
& \bnfalt & \enForce{\msg} & forcing
\\
& \bnfalt & \enLet{\var}{\expr_1}{\expr_2} & binding
\\
\fun & \bnfdef & \enFun[\pkg]{\var}{\expr} & functions
\\
\pkg & \bnfdef & \enApp \bnfalt \enFwk & packages
\\
\val \in \ValSet & \bnfdef & \var \bnfalt \enMe{} \bnfalt \cdots \bnfalt \enClosure{\fun}{\env} \bnfalt \addr \bnfalt \thunk \bnfalt \handle \bnfalt \enSkip
\end{grammar}
\end{mathpar}
\begin{mathpar}[\MathparNormalpar]
\text{variables} \quad \var \in \VarSet
\and
\text{addresses} \quad \addr \in \AddrSet
\and
\text{handles} \quad \handle \in \HandleSet
\and
\text{thunks} \quad \thunk \in \ThunkSet \bnfdef \enThunk{\enClosure{\fun}{\env}}{\val}
\and
\text{messages} \quad \msg \in \MsgSet \bnfdef \enMsg{\handle}{\thunk} 
\end{mathpar}
\begin{mathpar}
\begin{grammar}[@{}l]
stores &
\store
& \bnfdef & \cdot \bnfalt \store\extmap{\addr}{\val}
\\
message stores &
\eventmap, \callinmap
& \bnfdef & \cdot \bnfalt \eventmap\extMsg{\handle}{\thunk}
\\
continuations &
\cont 
& \bnfdef & \enSkipK \bnfalt \enLetK{\var}{\cont}{\expr}{\env}
\bnfalt \msg \bnfalt \enFrame{\msg}{\cont}
\\
states &
\state \in \StateSet & \bnfdef & \enState \bnfalt \enInitial 
\end{grammar}
\end{mathpar}
\caption{The syntax and the semantic domains of \semname, a core model of event-driven programs capturing \emph{enabledness} of events and \emph{allowedness} of invocations.}
\label{fig:model-syntax}
\end{figure}

\subsection{Thunks, Handles, and Messages}
\label{sec:messages}

The syntax of \semname is shown at the top of \figref{model-syntax}, which is a $\lambda$-calculus in a let-normal form.
The first line of expressions $\expr$ are standard, including variables and values $\val$,
control flow, and whatever base values and operations of interest (e.g., integers, tuples) but excluding function application.
We also assume that there are some heap operations of interest for manipulating a global store $\store$.
Events occur non-deterministically and return to the main event loop, so events must communicate through the shared, global heap.

The subsequent two lines split the standard call-by-value function application into multiple steps. The \enBind{\val_1}{\val_2} expression creates a thunk $\thunk$ by binding a function value $\enClosure{\fun}{\env}$ with an argument value $\val$. A thunk may be forced by direct invocation or indirect event dispatch.
Before a thunk may be forced, the \enAllow{\val} expression allocates a handle $\handle$ for a given thunk $\thunk$, which may be viewed as a \emph{permission} to force a particular thunk instance or \emph{message}. 
The permission to invoke may be revoked by \enDisallow{\val} on a handle $\handle$.
An \enInvoke{\val_1}{\val_2} term takes a handle-thunk pair $\handle\;\thunk$ to construct a message $\msg$ where $\handle$ must grant permission for forcing $\thunk$.
A message $\msg$ is simply a handle-thunk pair $\enMsg{\handle}{\thunk}$.

\newsavebox{\SBoxFunExecute}\sbox{\SBoxFunExecute}{$\fun_{\codej{|ci:x|}}$}
\newsavebox{\SBoxAddrTask}\sbox{\SBoxAddrTask}{$\addr_{\codej{t}}$}
\newsavebox{\SBoxFunTranslate}
\begin{lrbox}{\SBoxFunTranslate}\small
\begin{lstlisting}[language=Enable]
let k = bind #\usebox{\SBoxFunExecute}# #\usebox{\SBoxAddrTask}# in let h = allow k in invoke h k    ##
\end{lstlisting}
\end{lrbox}
As an example, let \usebox{\SBoxFunExecute} be the \codej{execute} method code and let \usebox{\SBoxAddrTask} be an address for an \codej{AsyncTask}. The standard function application $(\usebox{\SBoxFunExecute}\;\usebox{\SBoxAddrTask})$ would simply translate to
\[
\usebox{\SBoxFunTranslate}
\]
%
%
The reason to split function application into these steps is that thunks and handles are now first-class values and can be used in event dispatch.

The direct invocation expressions are mirrored with expressions for event dispatch. An \enEnable{\val} expression allocates a handle $\handle$ for a given thunk $\thunk$ and \emph{enables} it for the external event-processing system (i.e., gives the event-processing system permission to force the message $\enMsg{\handle}{\thunk}$), while the \enDisable{\val} expression \emph{disables} the message named by a handle $\handle$.
There is no expression directly parallel to \enInvoke{\val_1}{\val_2}. Instead, when an expression reduces to a value, control returns to the event-processing loop to (non-deterministically) select another enabled event.
The \enForce{\msg} expression is simply an intermediate in evaluation that represents a message that is forceable (i.e., has been permitted for forcing via \enkwAllow{} or \enkwEnable{}).


The remainder of the syntax is mostly standard, except that functions $\fun\colon \enFun[\pkg]{\var}{\expr}$ are tagged with a package $\pkg$, which we discuss subsequently in \secref{instrumented-semantics} regarding our instrumented semantics for recording traces.
The \enkwLet{} expression is variable binding with the usual semantics.
Values $\val$ of this expression language are variables $\var$, a variable used to bind the active event handle $\enMe{}$, whatever base values of interest $\cdots$, closures $\enClosure{\fun}{\env}$, store addresses $\addr$ (which are the values of heap-allocated objects), thunks $\thunk$, handles $\handle$, and unit $\enSkip$. Messages $\msg$ do not need to be first class, as they are referred to programmatically via their handles.

\begin{figure*}\footnotesize
\begin{mathpar}
\inferrule[Bind]{
}{
  \jstep{ \enState[\enBind{\fun}{\val}] }%
        { \enState[ \enThunk{\enClosure{\fun}{\env_1}}{\val} ] }
}
\;\;
\inferrule[Disable]{
}{
  \jstep{ \enState[\enDisable{\handle}][\env][\store][ \eventmap\extMsg{\handle}{\thunk} ] }%
        { \enState[ \enSkip ] }
}
\;\;
\inferrule[Disallow]{
}{
  \jstep{ \enState[\enDisallow{\handle}][\env][\store][ \eventmap ][ \callinmap\extMsg{\handle}{\thunk} ] }%
        { \enState[ \enSkip ] }
}
%

\inferrule[Enable]{
  \handle \notin \Domain(\eventmap)
}{
  \jstep{ \enState[\enEnable{\thunk}] }%
        { \enState[ \handle ][ \env ][ \store ][ \eventmap\extMsg{\handle}{\thunk} ] }
}
\;\;
\inferrule[Allow]{
  \handle \notin \Domain(\callinmap)
}{
  \jstep{ \enState[\enAllow{\thunk}] }%
        { \enState[ \handle ][ \env ][ \store ][ \eventmap ][ \callinmap\extMsg{\handle}{\thunk} ] }
}
\;\;
\inferrule[Invoke]{
  \enMsg{\handle}{\thunk} \in \callinmap
}{
  \jstep{ \enState[ \enInvoke{\handle}{\thunk} ] }%
        { \enState[ \enForce{\enMsg{\handle}{\thunk}} ] }
}

\inferrule[Force]{
  \enMsg{\handle}{\enThunk{\enClosure{ (\enFun[\pkg']{\var'}{\expr'}) }{\env'}}{\val'}} = \msg
}{
  \jstep{ \enState[ \enForce{\msg} ] }%
        { \enState[ \expr'\subst{\val'}{\var'}\subst{\handle}{\enMe} ][ \env'\extmap{\enMe}{\addr}\extmap{\var'}{\addr'} ]%
[ \store ]%
[ \eventmap ][ \callinmap ][ \enFrame{\msg}{\cont} ] }
}
\;\;
\inferrule[InvokeDisallowed]{
  \msg = \enMsg{\handle}{\thunk}
  \\
  \msg \notin \callinmap
}{
  \jstep%
        { \enState[ \enInvoke{\handle}{\thunk} ] }%
        { \enState[ \enSkip ][ \env ][ \store ][ \eventmap ][ \callinmap ][ \enSkipK ] }
}
\;\;
\inferrule[Event]{
  \enMsg{\handle}{\thunk} \in \eventmap
}{
  \jstep{ \enState[ \val ][ \env ][ \store ][ \eventmap ][ \callinmap ][ \enSkipK ] }%
        { \enState[ \enForce{\enMsg{\handle}{\thunk}} ][ \env ][ \store ][ \eventmap ][ \callinmap ][ \enMsg{\handle}{\thunk} ] }%
}
%
\end{mathpar}
\caption{Semantics. Create thunks and allocate handles to disable, disallow, enable, and allow messages.}
\label{fig:model-semantics-event}
\end{figure*}

\subsection{Semantics: Enabling Is Not Enqueuing}

In contrast to the Android implementation, the state of a $\semname$ program does not have a queue.
Rather, the external environment (e.g., user interactions) is modeled by the non-deterministic selection of an enabled event. When an event is processed, it begins execution in the current state, so the effects of external events can be modeled by effects on the global heap.
This model eliminates the complexity of a queue that is an implementation detail for the purposes of lifestate specification.

We consider an abstract machine model with a store and a continuation, and we enrich it with an \emph{enabled events} store $\eventmap$, which is a finite map from handles to thunks, and an \emph{allowed calls} store $\callinmap$, which is also a map from handles to thunks. These stores can also be seen as the set of enabled and allowed messages, respectively. And thus a machine state $\state\colon \enState$ consists of an expression $\expr$, a store $\store$, enabled events $\eventmap$, allowed calls $\callinmap$, and a continuation $\cont$ (shown at the bottom of \figref{model-syntax}).
The store $\store$ is a standard, finite map from addresses $a$ to values $v$.
We explain the special state \enkwInitial{} while discussing the trace semantics in \secref{instrumented-semantics}.
A continuation $\cont$ can be the top-level continuation $\enSkipK$ or a continuation for returning to the body of a $\enkwLet$ expression $\enLetK{\var}{\cont}{\expr}{\env}$, which are standard.
Continuations are also used to record the active messages via $\msg$
and $\enFrame{\msg}{\cont}$ corresponding to the run-time stack of
activation records.

We define an operational semantics in terms of the judgement form $\jstep{\state}{\state'}$ for a small-step transition relation. In \figref{model-semantics-event}, we show the inference rules defining the reduction steps related to creating, enabling-disabling, allowing-disallowing, and finally forcing messages.
The \TirName{Bind}, \TirName{Disable}, \TirName{Disallow}, \TirName{Enable}, and \TirName{Allow} rules follow closely the informal semantics discussed previously in \secref{messages}. Observe that \TirName{Enable} and \TirName{Allow} are parallel in that they both allocate a fresh handle $\handle$, and \TirName{Disable} and \TirName{Disallow} look up a message via its handle. The \emph{only} difference between \TirName{Enable} and \TirName{Disable} versus \TirName{Allow} and \TirName{Disallow} is that the former pair manipulate the enabled events $\eventmap$, while the latter touches the allowed calls $\callinmap$.
We write $\eventmap\extMsg{\handle}{\thunk}$ for the map that extends $\eventmap$ with a mapping $\enMsg{\handle}{\thunk}$ (i.e., is the same as $\eventmap$ except at $\handle$) and similarly for other maps.

The \TirName{Invoke} and \TirName{Event} rules have similar parallels. The \TirName{Event} rule says that when the expression is a value $\val$ and the continuation is the top-level continuation $\enSkipK$, then a message is a non-deterministically chosen from the enabled events $\eventmap$ to force. For instrumentation purposes (see \secref{instrumented-semantics}), we overload continuations to record the event message $\enMsg{\handle}{\thunk}$ in the continuation.
The \TirName{Invoke} rule checks that the given handle-thunk pair is an allowed message in $\callinmap$ before forcing.

The \TirName{Force} rule implements the ``actual application'' that reduces to the function body $\expr'$ with the argument $\val'$ substituted for the formal $\var'$. So that the forced message has ready access to its handle $\handle$, the $\handle$ is substituted for the \enMe{} parameter. Observe that an enabled event remains enabled after an \TirName{Event} reduction. A ``dequeuing'' event-processing semantics can be implemented by an event disabling itself (via \enDisable{\enMe}) on execution. To adequately model Android, it is important to be able to model both events that are self-disabling (e.g., the event that invokes \codej{onCreate}) and those that do not self-disable (e.g., the event that invokes \codej{onClick}).
Again for instrumentation purposes, we push the invocation message $\msg$ on the continuation (via $\enFrame{\msg}{\cont}$).
The \TirName{InvokeDisallowed} rule states that a disallowed message invocation terminates the running program by dropping the continuation.
The \TirName{Return} and \TirName{Finish} rules simply state that the recorded message $\msg$ frames are popped on return from a \TirName{Force} and \TirName{Event}, respectively.
We elide these rules and the remaining reduction rules, as they are as expected.

\subsection{Defining Callbacks and Observable Traces}
\label{sec:instrumented-semantics}

We wish to record traces of ``interesting'' transitions from which we mine lifestate specifications. Lifestate specifications abstract \TirName{Enable}, \TirName{Disable}, \TirName{Allow}, and \TirName{Disallow} transitions, which are unfortunately \emph{unobservable} or \emph{hidden}. That is, they are transitions in \semname that are not observable in an implementation like Android. Our challenge is to infer lifestate specifications from \emph{observable} transitions like, ``An event message $\msg$ was initiated.''

\begin{figure}[tb]\small
\begin{mathpar}
\begin{grammar}[@{}l]\\[-5ex]
transitions &
\trans \in \transset \subseteq \TransSet & \bnfdef & \enObs{\obsty}{\msg}  \bnfalt \silentTrans \bnfalt \enInit
\\
observable kinds &
\obsty & \bnfdef & \enkwEvt \bnfalt \enkwDis \bnfalt \enkwCb \bnfalt \enkwCi \bnfalt \enkwRet 
\\[1ex]
traces &
\trace \in \seqof{\TransSet} & \bnfdef & \cdot \bnfalt \trans \trace 
\end{grammar}
\end{mathpar}
\footnotesize\begin{mathpar}
\inferrule[ForceCallback]{
  \enMsg{\handle}{\enThunk{\enClosure{ (\enFun[\enAppSub]{\var'}{\expr'}) }{\env'}}{\val'}} = \msg
  \\
  \enFwk = \Package(\cont)
}{
  \jstepins[\enCb{\msg}]
        { \enState[ \enForce{\msg} ] }%
        { \enState[ \expr'\subst{\val'}{\var'}\subst{\handle}{\enMe} ][ \env'\extmap{\enMe}{\addr}\extmap{\var'}{\addr'} ]%
[ \store ]%
[ \eventmap ][ \callinmap ][ \enFrame{\msg}{\cont} ] }
}

\inferrule[ForceCallin]{
  \enMsg{\handle}{\enThunk{\enClosure{ (\enFun[\enFwkSub]{\var'}{\expr'}) }{\env'}}{\val'}} = \msg
  \\
  \enApp = \Package(\cont)
}{
  \jstepins[\enCi{\msg}]
        { \enState[ \enForce{\msg} ] }%
        { \enState[ \expr'\subst{\val'}{\var'}\subst{\handle}{\enMe} ][ \env'\extmap{\enMe}{\addr}\extmap{\var'}{\addr'} ]%
[ \store ]%
[ \eventmap ][ \callinmap ][ \enFrame{\msg}{\cont} ] }
}
\end{mathpar}
\small\begin{mathpar}
\begin{array}{@{}r@{\;}c@{\;}l@{}}
\\[-0.5ex]
\Msg(\enLetK{\var}{\cont}{\expr}{\env}) & \defeq & \Msg(\cont) \\
\Msg(\msg) \defeq \Msg(\enFrame{\msg}{\cont}) & \defeq & \msg \\
\Package(\cont) & \defeq & \pkg \quad\text{if $\enMsg{\handle}{\enThunk{\enClosure{ (\enFun[\pkg]{\var}{\expr}) }{\env}}{\val}} = \Msg(\cont)$} \\
\end{array}
\end{mathpar}
\caption{Callbacks and callins are transitions between framework and app code.}
\label{fig:model-semantics-instrumented}
\end{figure}

In \figref{model-semantics-instrumented}, we define the judgment form $\jstepins{\state}{\state'}$, which instruments our small-step transition relation from the previous subsection with transition labels $\trans$.
%
Interesting observable transitions correspond primarily to introduction and elimination operations on messages.
We instrument the \TirName{Event} rule to record that event message $\msg$ was initiated (via observable $\enEvt{\msg}$).
Similarly, we instrument the \TirName{InvokeDisallowed} rule to record that an invocation of a message was disallowed (via $\enDis{\enMsg{\handle}{\thunk}}$).
We use $\silentTrans$ to instrument an uninteresting, ``don't care'' transition or an unobservable, ``don't know'' transition.
The allowed invocation of a message with the \TirName{Invoke} rule is an uninteresting transition, so it is simply labeled with $\silentTrans$.
These cases are simply adding a transition label to the rules from \figref{model-semantics-event}, so they are not shown here. 

Recall from \secref{overview} that we define a callback as an invocation that transitions from framework to app code and a callin as an invocation from app to framework code.
In $\semname$, this definition is captured crisply by the context in which a message is forced.
In particular, we say that a message is a \emph{callback} if the underlying callee function is an app function (package \enkwApp{}) and it is called from a framework function (package \enkwFwk{}) as shown in \TirName{ForceCallback}.
The $\Msg(\cdot)$ function inspects the continuation for the running, caller message.
The $\Package(\cdot)$ function gets the package of the running message.
Analogously, a message is a \emph{callin} if the callee function is in the \enkwFwk{} package, and the caller message is in the \enkwApp{} package (rule \TirName{ForceCallin}).
There is a remaining rule not shown here, \enkwForce{} (\TirName{ForceInternal}), where there is no switch in packages (i.e., $\pkg' = \Package(\cont)$ where $\pkg'$ is the package of the callee message). It is not an interesting transition, so it is labeled with $\silentTrans$.

These three cases for \enkwForce{} replace the \TirName{Force} rule from \figref{model-semantics-event}.
%
We also instrument the \TirName{Return} rule to record returning via $\enRet{\msg}$ to retain a call tree structure in the sequence of transitions.
Finally, to simplify subsequent definitions, we introduce a ``dummy'' \enkwInitial{} state and a ``dummy'' \enkwInit{} transition with an \TirName{Init} rule.
%
%
Any remaining rules defining the original transition relation $\jstep{\state}{\state'}$ not discussed here are simply labeled with the ``don't care'' transition label $\silentTrans$.



We write $\denote{\expr}$ for the path semantics of \semname{} expressions $\expr$ that collects the finite (but unbounded) sequences of alternating state-transition-state $\state\trans\state'$ triples according to the instrumented transition relation $\jstepins[\trans]{\state}{\state'}$.
From paths, we derive \emph{traces} $\trace$ in an expected manner by keeping transitions but dropping intermediate states.

%

%

The design of the trace recording in \toolname{} follows this instrumented semantics $\jstepins[\trans]{\state}{\state'}$ to obtain traces like the one shown in \figref{feedremover-trace}.
In particular, it maintains a message stack corresponding to the continuation $\cont$ to emit callback $\enkwCb$ and callin $\enkwCi$ transitions.


\section{Lifestate Specification: Abstracting Enabledness and Allowedness}
\label{sec:specification}

In this section, we define formally the meaning of lifestate rules building on the $\semname$ concrete model of event-driven programs from \secref{concrete}.

We consider the abstraction of messages to be a parameter of our approach.
We call the message abstraction of interest a message signature $\absmsg \in \absofset{\MsgSet}$ and assume it comes equipped with a standard concretization function $\conc : \absofset{\MsgSet} \rightarrow \powerset(\MsgSet)$ to give signatures meaning and an element-wise abstraction function $\abselem : \MsgSet \rightarrow \absofset{\MsgSet}$ to lift a message to its signature satisfying the soundness and best-abstraction relationship.
In the case of an Android method call like \codej{(t: AsyncTask).execute()} (which corresponds to a message
$\msg\colon \enMsg{\handle_{\codej{t}}}{\enThunk{\usebox{\SBoxFunExecute}}{\usebox{\SBoxAddrTask}}}$ for some handle $\handle_{\codej{t}}$, function \usebox{\SBoxFunExecute}, and object address \usebox{\SBoxAddrTask}), the message signature can be the Java method signature \codej{AsyncTask.execute()}.
In \toolname{}, we make a further refinement on Java method signatures to split on values of primitive type, so \codej{Button.setEnabled(false)} and \codej{Button.setEnabled(true)} are considered distinct message signatures.
For the implementation, the element-wise abstraction function $\abselem$ is applied to recorded messages to produce message signatures for lifestate mining.

\begin{figure}\centering\small
\begin{mathpar}
\begin{grammar}[@{}l]
lifestate specs &
\absof{\directiveset} \in \SpecSet & \bnfdef & \cdot \bnfalt \absof{\directiveset}\ext{\absof{\directive}}
\end{grammar}
\end{mathpar}
\begin{mathpar}
\begin{grammar}[@{}l]
rules &
\absof{\directive} & \bnfdef &
\absforce\dashrightarrow\absmsg
\quad
\dashrightarrow
\bnfdef
\specArrow\specEnableOp \bnfalt \nspecArrow\specDisableOp \bnfalt
\specArrow\specAllowOp \bnfalt \nspecArrow\specDisallowOp
\\
forcings &
\absforce & \bnfdef & \absmsg \bnfalt \initforce
\end{grammar}
\end{mathpar}
\begin{mathpar}
\begin{grammar}[@{}l]
signature transitions &
\absof{\trans} \in \absTransSet & \bnfdef & \enObs{\absof{\obsty}}{\absmsg} \bnfalt \enInit
\\
&
\absof{\obsty} & \bnfdef & \enkwEvt \bnfalt \enkwCi \bnfalt \enkwDis
\\[1ex]
signature traces &
\abstrace \in \abstraceset \subseteq \absTransSeqSet & \bnfdef & \cdot \bnfalt \absof{\trans} \absof{\trace}
\\[1ex]
signature stores &
\absof{\eventmap}, \absof{\callinmap}
& \bnfdef & \cdot \bnfalt \absof{\eventmap}\ext{\absmsg}
\\
signature states &
\absof{\state} \in \absofset{\StateSet} & \bnfdef & \enSigState
\end{grammar}
\end{mathpar}
\footnotesize\begin{mathpar}
\inferrule[Enabled]{
  \absmsg \in \absof{\eventmap}
  \\
  \absof{\eventmap'} = 
  \absof{\eventmap}
  \union \allEnables{\absmsg}
  \\
  \absof{\callinmap'} =
  \absof{\callinmap}
  \union \allAllows{\absmsg}
}{
  \jstep{
    \enSigState[ (\enEvt{\absmsg})\absof{\trace}  ]
  }%
  {
    \enSigState[ \absof{\trace} ][ \absof{\eventmap'} - \allDisables{\absmsg} ][ \absof{\callinmap'} - \allDisallows{\absmsg} ]
  }
}

\inferrule[Allowed]{
  \absmsg \in \absof{\callinmap}
  \\
  \absof{\eventmap'} = 
  \absof{\eventmap}
  \union \allEnables{\absmsg}
  \\
  \absof{\callinmap'} =
  \absof{\callinmap}
  \union \allAllows{\absmsg}
}{
  \jstep{
    \enSigState[ (\enCi{\absmsg})\absof{\trace}  ]
  }%
  {
    \enSigState[ \absof{\trace} ][ \absof{\eventmap'} - \allDisables{\absmsg} ][ \absof{\callinmap'} - \allDisallows{\absmsg} ]
  }
}

\inferrule[Disallowed]{
  \absmsg \notin \absof{\callinmap}
}{
  \jstep{
    \enSigState[ (\enDis{\absmsg})\absof{\trace}  ]
  }%
  {
    \enSigState[ \absof{\trace} ]
  }
}

\inferrule[Initialized]{
  \absof{\eventmap} = \allEnables{\initforce}
  \\
  \absof{\callinmap} = \allAllows{\initforce} 
}{
  \jstep{
    \enSigState[ (\enInit)\absof{\trace} ][ \cdot ][ \cdot ]
  }%
  {
    \enSigState
  }
}
\end{mathpar}
\begin{mathpar}
\\
\begin{array}{r@{}c@{}l@{\quad}r@{}c@{}l}
\allEnables{\absforce} & \defeq & \SetST{ \absmsg }{ \specEnable{\absforce}{\absmsg} \in \absof{\directiveset} }
&
\allDisables{\absforce} & \defeq & \SetST{ \absmsg }{ \specDisable{\absforce}{\absmsg} \in \absof{\directiveset} }
\\[1ex]
\allAllows{\absforce} & \defeq & \SetST{ \absmsg }{ \specAllow{\absforce}{\absmsg} \in \absof{\directiveset} }
&
\allDisallows{\absforce} & \defeq & \SetST{ \absmsg }{ \specDisallow{\absforce}{\absmsg} \in \absof{\directiveset} }
\\[2ex]
\end{array}
\end{mathpar}
\rule{0.75\linewidth}{0.15pt}
\begin{mathpar}
\begin{array}{rcl}
\denote{\cdot} & : & \SpecSet \rightarrow
                     \powerset( \seqof{\absofset{\TransSet}} )
\\
\denote{\absof{\directiveset}} & \defeq &
\SetST{ \absof{\trace} }{
  \enSigState[ \absof{\trace} ][ \cdot ][ \cdot ]
  \longrightarrow_{\absof{\directiveset}}^{\ast}
  \enSigState[ \cdot ]
  \quad\text{for some $\absof{\eventmap}, \absof{\callinmap}$}
}
\end{array}
\end{mathpar}
\caption{Meaning of lifestate specifications. A lifestate rule $\absof{\directive}$ is an abstraction of
enabling $\enEnable{\val}$, disabling $\enDisable{\val}$, allowing $\enAllow{\val}$, or disallowing $\enDisallow{\val}$ transitions.}
\label{fig:specification}
\end{figure}

A lifestate specification $\absrules$ is a set of rules where a rule $\absrule$ is an enable $\specArrow\specEnableOp$, disable $\nspecArrow\specDisableOp$, allow $\specArrow\specAllowOp$, or disallow $\nspecArrow\specDisallowOp$ constraint.
We define the meaning of lifestate specifications by defining an abstraction of the concrete transition relation $\jstep{\state}{\state'}$ for $\semname$.

To do so, we consider signature transitions $\absof{\trans}$ and a signature traces $\absof{\trace}$.
A signature transition $\absof{\trans}$ are simply analogous to concrete transitions $\trans$, except over signatures $\absmsg$ instead of concrete messages $\msg$.
As input to specification mining, we are only concerned with the event $\enkwEvt$, callin $\enkwCi$, and disallowed $\enkwDis$ transition kinds, so we drop the callback $\enkwCb$ and return $\enkwRet$ kinds.
A signature trace $\absof{\trace}$ is simply a sequence of signature transitions.
To define the meaning of signature transitions and traces, we lift the concretization on message signatures to signature transitions $\absof{\trans}$ and traces $\absof{\trace}$ in the expected way.

We can then define a signature store $\absof\eventmap, \absof\callinmap$ as a set of message signatures, which is an abstract analogue of a message store.
In a signature state $\absof{\state}\colon \enSigState$, the signature store $\absof\eventmap$ is a set of enabled message signatures (i.e., abstract messages) and $\absof\callinmap$ is a set of allowed message signatures.
For convenience, we refer to the union $\absof\eventmap \union \absof\callinmap$ of enabled and allowed message signatures in a state as the set of \emph{permitted} signatures and all other message signatures that are disabled or disallowed as the \emph{prohibited} signatures.

A signature state $\absof{\state}\colon \enSigState$ also includes a signature trace $\absof{\trace}$ that controls the execution of a corresponding abstract machine defined by the transition relation $\jstep[\absrules]{\absof{\state}}{\absof{\state}'}$ that is parametrized by a set of rules $\absrules$ and defined in \figref{specification}.
This transition relation is relatively straightforward: it considers the signature transition $\absof{\trans}$ at the beginning of the $\absof{\trace}$ and determines whether $\absof{\trans}$ is permitted according to the set of permitted signatures.
If it is permitted, it updates the signature stores $\absof{\eventmap}$ and $\absof{\callinmap}$ according to the lifestate rules $\absrules$.
For example, the \TirName{Enabled} rule checks if the current transition $\enEvt{\absmsg}$ can happen by checking if $\absmsg$ is in the enabled set $\absof{\eventmap}$.
If so, it updates the permitted state according to the rules $\absrules$. The helper function $\allEnables{\absforce}$ gathers the set of messages that is enabled by $\absforce$ and similarly for disable, allow, and disallow rules.
For presentation cleanliness, we leave the rules $\absrules$ implicit in the inference rules since it is constant.

Finally, the meaning of a lifestate specification $\denote{\absrules}$ is the set of signature traces that do not get stuck during reduction (where $\longrightarrow_{\absof{\directiveset}}^{\ast}$ is the reflexive-transitive closure of the single step relation).
And this definition gives rise to a natural description for when a trace is described by or sound with respect to a specification:
\begin{definition}[Trace Soundness]\label{def:trace-soundness}
A signature trace $\absof{\trace}$ is \emph{sound} with respect to
a lifestate specification $\absof{\directiveset}$ iff
$\absof{\trace}$ is described by $\absof{\directiveset}$ (i.e.,
$\absof{\trace} \in \denote{\absof{\directiveset}}$).
Or in other words, the signature trace $\absof{\trace}$ can be reduced
without getting stuck (i.e., $\enSigState[\absof{\trace}][\cdot][\cdot] \longrightarrow_{\absof{\directiveset}}^{\ast}
\enSigState[\cdot]$ for some $\absof{\eventmap}, \absof{\callinmap}$).
\end{definition}

A specification is then sound with respect to a $\semname$ program if its signature traces abstract the set of concrete traces of the program:
\begin{definition}[Specification Soundness]\label{def:specification-soundness}
A lifestate specification $\absof{\directiveset}$ is \emph{sound} with
respect to an expression $\expr$ iff for every trace $\trace$ of
$\expr$, it is in the concretization of a signature trace $\absof{\trace}$
of $\absof{\directiveset}$ (i.e., formally, $\denote{\expr} \subseteq
\conc(\denote{\absof{\directiveset}})$).
\end{definition}

\section{Trace Slicing for Multi-Object Event-Driven Protocols}
\label{sec:slicing}


As the specification mining problem is severely under-constrained,
we wish to constrain the search space of specifications as much as possible before applying the learning algorithms.
As a first step,
we wish to slice a trace into sub-traces that group together related messages or method invocations that (likely) correspond to different protocols.
A well-known issue is that related method invocations may involve multiple objects.
This issue is exacerbated in event-driven protocols where event messages correspond to some internal data structures of the framework rather than method calls per se.
For instance, in our running example, a call to the callin \codej{(b:Button).setEnabled($\ldots$)} that enables or disables a button \codej{b} should be associated with a \fmtevt{Click} event for \codej{b}.

One reasonable and common assumption is that two related messages should use some common values that are readily accessible, such as arguments for method invocations.
We call this heuristic, the \emph{argument-sharing strategy}, and leave as a parameter of the strategy a function $\Args : \MsgSet \rightarrow \finitepowerset(\ValSet)$ that specifies the arguments of a message.
This heuristic is similar to the one employed by \citet{DBLP:conf/kbse/PradelG09} for mining multi-object API protocols.
Following the argument-sharing strategy, we can slice observations so that they contain only observable transitions that share a common argument (this strategy is essentially described as trace slicing~\cite{DBLP:conf/tacas/ChenR09}). 

However, one issue in trace slicing for event-driven object protocols is that the messages of interest are the event messages (in $\enEvt{\msg}$) and the callin messages (in $\enCi{\msg'}$).
As noted above, the ``argument'' of interest for the \fmtevt{Click} event message is the button \codej{b} buried inside internal implementation-specific data structures.

From the \codej{(b:Button).setEnabled($\ldots$)} callin and \fmtevt{Click} event example, we also see that regardless of how \codej{b} may be stored in the \fmtevt{Click} event message, the \fmtevt{Click} event eventually invokes the callback \codej{$l$.onClick($b$)} on some listener \codej{l} where button \codej{b} becomes an argument (retrieving \codej{b} in some implementation-specific manner).
Thus we extend the argument-sharing strategy to event messages by defining the arguments of an event message $\msg$ as the arguments of any callbacks that it invokes (instead of $\msg$'s own arguments).
%
Once the traces have been sliced for each object, we apply the message abstraction function $\abselem$ and partition the traces based on the abstraction of the sliced object.

While our particular argument-sharing strategy focuses on associating callbacks with their initiating events, this strategy is quite flexible in that other appropriate context information (e.g., values reached through the heap) could be incorporated into the trace slice.

\section{Mining Lifestate Specifications}\label{sec:mining}
%
In the previous sections, we formalized a model for event-driven systems like Android centered around enabled events and allowed callins. From this model, we derive the notion of lifestate specifications. In this section, we discuss an application of specification mining techniques to learn lifestate specifications.
%
For a set of signature traces $\abstraceset$,
let $\Alphabet$ be the set containing all the abstract messages
contained in the traces $\abstraceset$ and $\initforce$.
We learn specifications where the messages in the rules
are elements of $\Alphabet$, hence restricting our learning
algorithms to an abstraction of messages that we observed concretely.
We denote with $\RuleSet$ the set of all the possible rules
that can be learned from $\Alphabet$.

The goal of the mining algorithms is to find a specification
$\absof{\directiveset} \in \powerset(\RuleSet)$ and to this end we consider 
3 different approaches. Two of these are applications of an off-the-shelf 
machine learning tool Treba~\cite{hulden2012treba} learning probabilistic models of 
the interaction between the app and the Android framework, and a 
symbolic sampling one based on model counting of propositional Boolean
formulas~\cite{sharpsat}, implemented with pySMT~\cite{pysmt} and the solver sharpSAT~\cite{sharpsattool}.



\JEDI{
  Better definition of the problem, (state space)
  \begin{itemize}
  \item Define the set of possible rules from the alphabet
  \item What is $\powerset(\SpecSet)$    
  \end{itemize}
}

\paragraph{Probabilistic Models.}
Hidden Markov Models (HM\-M) and Probabilistic Finite State Automata (\pfsa) are 
common statistical models famed for their simplicity and accuracy in modeling complex,
often partially observable processes.
%
%
The challenge of learning a probabilistic model of the interaction between the 
Android framework and an app over the alphabet of signature messages 
is that in general neither model has a direct interpretation.
For a state $x_i$ of an 
a HMM or PFSA $\mathcal{A}$, we define the set of permitted message signatures 
to be the labels of all enabled outgoing transitions 
and the set of prohibited message signatures as its complement. 

For a state
$x_i$
and an enabled outgoing transition labeled $\absmsg$, we learn the lifestate enable (or resp. allow) rules that $\absmsg$ permits 
all message signatures prohibited in $x_i$ but permitted in the 
target set of this transition.
Similarly, for the state $x_i$ and an enabled outgoing transition labeled $\absmsg$, we learn  disable (or disallow) rules that $\absmsg$ prohibits 
all messages permitted in $x_i$ that become prohibited along the transition.


The final lifestate specification $\absof{\directiveset}$ is the union of permit 
and prohibit rules of all states of the automaton.
To identify high probability rules, we associate with each rule 
$\absmsg_i\dashrightarrow\absmsg_j \in \absof{\directiveset}$ a weight
that is the sum the probabilities of all transition that define the rule normalized by the 
number of permitted sets $\absmsg_i$ is in.

\paragraph{Symbolic Sampling via \sharpsat.}
\input{sharpsat}



\section{Empirical Evaluation: Mining}
\label{sec:evaluation}

Here, we evaluate empirically whether our process finds meaningful lifestate specifications, as well as compare the quality of the specifications obtained from different mining algorithms.

\paragraph{Research Questions.}
Given a candidate specification (i.e., a set of lifestate rules) obtained from a learning method, we consider the following research questions.
\begin{itemize}\itemsep 0pt
\item
\JEDI{We measure sufficiency by computing recall for traces labeled by soundness.  We cannot directly observe precision.}
\emph{RQ1 (Sufficiency): Is a mined specification sound with respect to previously unseen traces?} We say that a specification is sufficient for explaining a set of traces if each trace is sound with respect to the specification (according to Definition 1 in section 4). Foremost, we seek specifications that are, in the limit, sufficient to explain all actual Android behavior.
\item
\JEDI{We measure veracity of a spec by computing precision for rules labeled by actual Android behavior}
\emph{RQ2 (Veracity): For each rule in a mined specification, does the rule correspond to actual Android behavior?} We say that a specification is veracious if it captures when a message should be enabled or allowed during the execution of an app and when it should not. 
In the end, we want to find rules that in fact describe the true enabling, disabling, allowing, and disallowing behavior of Android.
%

\end{itemize}

\paragraph{Experimental Methodology.}

We consider the mining algorithms from \secref{mining} based on probabilistic models (\hmm and \pfsa) and symbolic sampling (\sharpsat).
For the symbolic sampling approach, we use two weight thresholds 0.6 and 0.8, corresponding intuitively to keeping only the rules that explain slightly more than half the paths and those that explain most paths, respectively.
This selection of parameters gives us four learning methods to evaluate that we abbreviate as \sharpsat-0.6, \sharpsat-0.8, \hmm, and \pfsa.

Traces recorded by \toolname{} are sliced for each object, abstracted into message signatures, and grouped by the framework type of the slicing object.
Thus, we have a corpus of sliced and abstracted traces for each framework type of interest.
To evaluate a learning method, we divide a trace corpus into a training set and a testing set using 5-fold cross validation.

\newcommand{\tblunitstyle}{\relsize{-1}}
\newcommand{\fmtbench}[1]{{\ttfamily\fontseries{l}\selectfont\fwktystyle\color{black}#1}}
\newcommand{\AsyncTaskBench}{\fmtbench{AsyncTask}}
\newcommand{\FragmentBench}{\fmtbench{Fragment}}
\newcommand{\FragmentFourBench}{\fmtbench{FragmentV4}}
\newcommand{\ButtonBench}{\fmtbench{Button}}
\newcommand{\fwktype}{f-type}

\paragraph{Corpus: Generating Sliced and Abstracted Traces}

We begin with 133 traces of running Android applications recorded by \toolname{}.
These traces were generated from a corpus of 1909 apps retrieved from Github.

We used two different methods for generating traces: manual exercising of the app and automatic by means of the
UI Exerciser Monkey.
We chose to generate some traces using manual exercising because apps require user logins, API keys, and other inputs that present problems for automatic exploration tools
(e.g., \cite{DBLP:conf/sigsoft/MachiryTN13,DBLP:conf/oopsla/AzimN13,DBLP:conf/kbse/AmalfitanoFTCM12}).
The manual method resulted in 65 traces and the automatic in 68.
The process for manual trace generation was to exercise the app in a normal manner for 10 minutes under instrumentation. For the automatic Monkey, a script loaded the application on an Android emulator, waited 3 minutes for any application initialization, and then pulsed the UI exercising 40 times. Each time inputting 50 random events and delaying 30 seconds in between.

\begin{table}\small
\caption{The data set is partitioned by framework type (\fwktype) on the sliced object. For each trace set, we give the number of traces, the average length of the traces (len), the total number of distinct event message signatures (\enkwEvt{}s), callin signatures (\enkwCi{}s), and the size of the specification space ($\card{\SpecSet}$).}
\label{tbl:objects}
	\begin{tabular*}{\linewidth}{@{\extracolsep{\fill}} l r r r r r @{}}\\\toprule
	       & traces & len & \enkwEvt{}s & \enkwCi{}s & $\card{\SpecSet}$ \\
	\fwktype & \tblunitstyle (num) & \tblunitstyle (mean)  & \tblunitstyle (num)& \tblunitstyle (num) & \tblunitstyle (num) \\
	\midrule
	\FragmentBench & 52 & 9.7 & 6 & 19 & 1275\\
	\FragmentFourBench & 124 & 9.0 & 10 & 34 & 3916\\
	\ButtonBench & 654 & 4.6 & 4 & 28 & 2080\\
	\AsyncTaskBench & 63 & 3.8 & 1 & 4 & 55 \\
	\midrule
	summary & 893 & 6.8 & 21 & 85 & 7326 \\
	\bottomrule	
	\end{tabular*}
\end{table}

From the corpus of 133 recorded traces, we apply callback-driven trace slicing from \secref{slicing} to get 6134 sliced traces over 184 framework types.
Each framework type corresponds to a partition of the 6134 sliced traces.
We selected four of these framework types, which had trace sets of sufficient length and diversity of events and callins, for evaluating our learning methods.
In \tblref{objects}, we list some statistics about the data set. The specification space ($\card{\SpecSet}$) is the number of possible lifestate rules as determined by the number of observed events (\enkwEvt{}s) and callins (\enkwCi{}s).
The summary line is the total number of traces, the mean of mean length of traces, the total number of event and callin signatures, the total specification space.


\paragraph{Candidate Specifications.}

In \tblref{specifications}, we show the number of each rule kind learned using each method on the first training set.
Our first observation is that the probabilistic model-based methods generate many more candidate rules than the symbolic sampling-based method.
This observation is not unexpected, as the symbolic sampling-based method takes a more conservative approach for generating rules: it generates rules based on the frequency of explainable signature traces.
Our second observation is that event-disables are the least frequently derived rules.
This observation is also not unexpected, as event-disables are the least constrained rule kind.
The total line is the total number of rules of each kind. The magnitude is not meaningful, but it shows the relative frequency of the rule kinds.

\begin{table}\small
\caption{Mined candidate specifications.
For each trace set for a framework type, we apply each method to learn on a training set (using 5-fold cross validation).
This table shows the number of lifestate rules of each kind on the first training set.
}
	\label{tbl:specifications}
    \begin{tabular*}{\linewidth}{@{\extracolsep{\fill}} l l r r r r @{}}\\\toprule
    & &  \multicolumn{4}{c}{lifestate rules} \\ \cmidrule{3-6}
    & & $\specArrow\specEnableOp$ & $\nspecArrow\specDisableOp$ & $\specArrow\specAllowOp$ & $\nspecArrow\specDisallowOp$ \\
    \fwktype & method & \tblunitstyle (num) & \tblunitstyle (num) & \tblunitstyle (num) & \tblunitstyle (num) \\
    \midrule
    \FragmentBench		& \sharpsat{}-0.6 & 2 & 0 & 17 & 5\\
	\FragmentBench		& \sharpsat{}-0.8 & 0 & 0 &  3 & 0\\
	\FragmentBench		& \hmm	& 7 & 12 & 48 &	32\\
	\FragmentBench		& \pfsa	& 12 & 14 & 	39 & 41\\
	\FragmentFourBench	& \sharpsat{}-0.6 & 2 & 0 & 24 & 2\\
	\FragmentFourBench	& \sharpsat{}-0.8 & 0 & 0 & 2 & 0\\
	\FragmentFourBench	& \hmm	& 39 & 24 & 235 & 209\\
	\FragmentFourBench	& \pfsa	& 27 & 45 & 145 & 213\\
	\ButtonBench		& \sharpsat{}-0.6 & 7 & 1 & 30 & 7\\
	\ButtonBench		& \sharpsat{}-0.8 & 2 & 1 & 7 & 4\\
	\ButtonBench		& \hmm	& 29 & 22 & 87 & 89\\
	\ButtonBench		& \pfsa	& 20 & 25 & 87 & 90\\
	\AsyncTaskBench		& \sharpsat{}-0.6 & 1 & 0 & 2 & 0\\
	\AsyncTaskBench		& \sharpsat{}-0.8 & 0 & 0 & 1 & 0\\
	\AsyncTaskBench		& \hmm	& 2 & 4 & 6 & 8\\
	\AsyncTaskBench		& \pfsa	& 1 & 2 & 5 & 5\\
    \midrule
    total & & 151 & 150 & 738 & 705\\
    \bottomrule
    \end{tabular*}
%
\end{table}

\paragraph{RQ1: Measuring Sufficiency.}

Given a testing set and a candidate specification, we measure sufficiency as the ratio of traces that are sound with respect to the specification to the number of traces in the testing set.

\begin{figure}\centering
  \includegraphics[width=\linewidth]{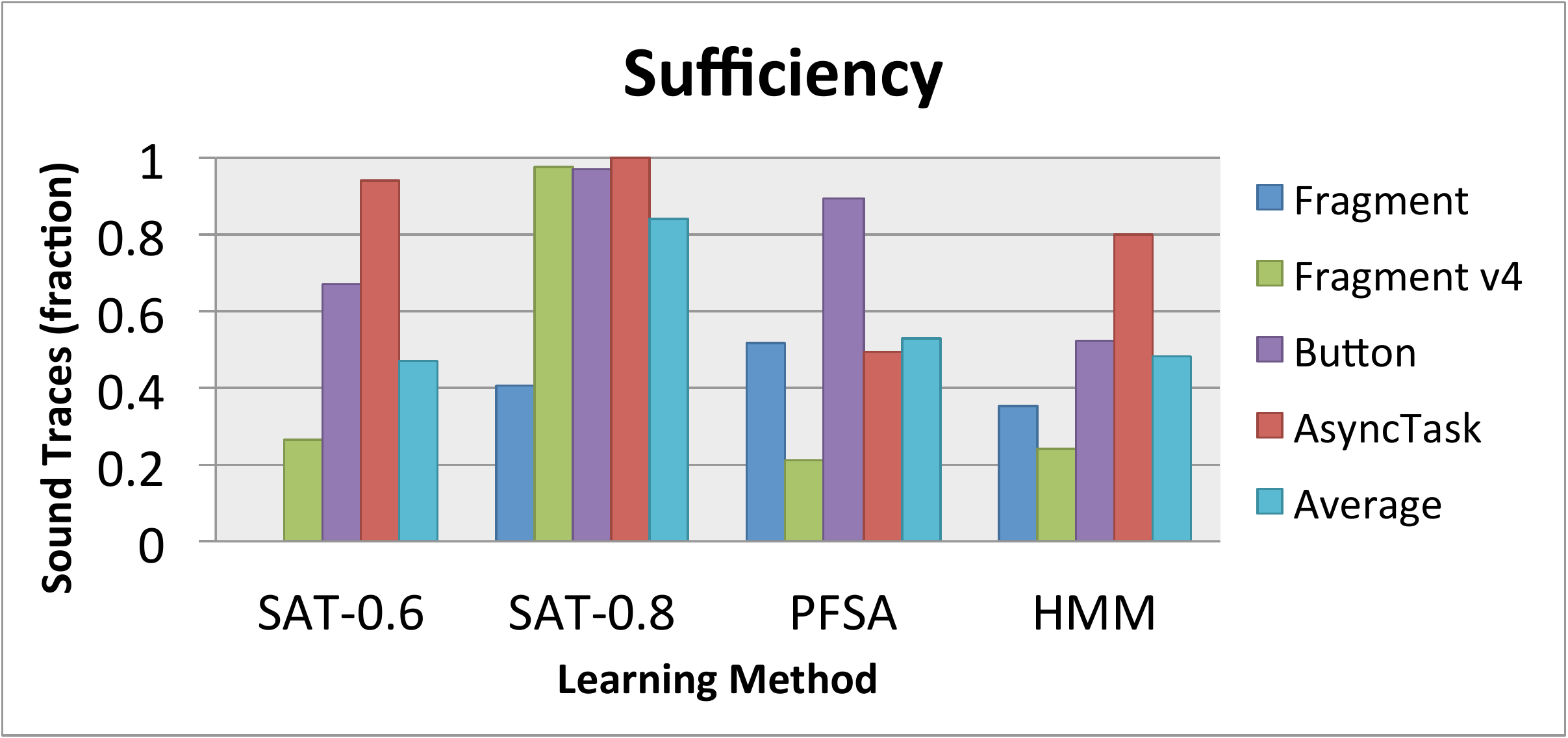}
\caption{Sufficiency measures the fraction of sound traces of a given testing set for a given specification (higher is better).
For each training set for a framework type, we measure sufficiency for each learned specification on the corresponding testing set.
The y-axis shows the average sufficiency measure over the testing sets (from 5-fold cross validation) for each framework type grouped by learning method. The rightmost bar in each group shows the average over the four framework types.}
 \label{fig:soundnessres}
\end{figure}

\Figref{soundnessres} shows the sufficiency measure for each learned specification, grouped by method. No one method dominates the others by this measure. For example, the \pfsa{} method is the only one that generates rules that have soundness fraction over 0.5 for \FragmentBench{} but is worse than all other methods for \AsyncTaskBench{}.
The sufficiency measure for \sharpsat-0.8 is generally high, but recall from \tblref{specifications} that \sharpsat-0.8 learns far fewer rules than the other methods. It is quite conservative in the rules it learns, but for the rules that it does learn, they tend to generalize well.
The \sharpsat-0.6 and \pfsa methods appear to balance producing more candidate rules and ones that generalize reasonably well. We thus consider manual triage of the rules produced by these two methods for evaluating veracity.

\paragraph{RQ2: Evaluating Veracity.}

\newcommand{\veraactual}{actual}
\newcommand{\veraimplied}{implied}
\newcommand{\veratrueimplied}{not false}
\newcommand{\verafalse}{false}

To evaluate veracity, we take a candidate lifestate specification and manually triage the rules and categorize them with respect to the correctness in capturing the true behavior of the Android framework.
We consider the following categories of correctness:
\begin{inparadesc}
\item[(actual)] the rule abstracts understood Android behavior from reading Android documentation or the framework code;
\item[(direct)] the rule is actual \emph{and} corresponds to the root cause in Android (e.g., a learned enable rule corresponds to direct enabling in Android);
\item[(false or unknown)] the rule appears to contradict understood Android rules.
\end{inparadesc}
If we cannot be sure that a learned rule corresponds to a \veraactual{} rule, then we conservatively classify it as \verafalse{}.

\newcommand{\p}{\phantom{1}}
\begin{table}\small
\caption{Veracity is a measure of precision in finding actual rules by manually triaging the rules found by each learning method.
The actual column shows the number of learned rules that correspond understood Android behavior. The direct shows the number of actual rules that also correspond to the root enable, disable, allow, or disallow in Android.
}
\label{tbl:ManualTriage}
\begin{tabular*}{\linewidth}{@{\extracolsep{\fill}} l l r r r @{}}\\\toprule
		         &          & rules & actual & direct \\
		\fwktype & method   & \tblunitstyle num & \tblunitstyle num & \tblunitstyle num (frac) \\
		\hline
		\FragmentBench    & \sharpsat-0.6 & 20 & 10 & 7 (0.7)\\ %
		\FragmentBench    & \pfsa     & 20 & 14 & 11 (0.8)  \\ %
		\FragmentBench    & \hmm & 20 & 5  & 4 (0.8) \\ %
		\FragmentBench    & \text{total unique} & 42 & 22 & 18 (0.8) \\\midrule %
		\FragmentFourBench & \sharpsat-0.6 & 20 & 8 & 7 (0.9)  \\ %
		\FragmentFourBench & \pfsa     & 20 & 12 & 10 (0.8) \\ %
		\FragmentFourBench & \hmm & 19 & 11  & 9 (0.8) \\ %
		\FragmentFourBench & \text{total unique} & 58 & 24 & 19 (0.8) \\\midrule %
		\ButtonBench      & \sharpsat-0.6 & 20 & 6 & 3 (0.5) \\ %
		\ButtonBench      & \pfsa  & 20 & 7 & 7 (1\phantom{.0}) \\ %
		\ButtonBench      & \hmm & 20 & 3  & 3 (1\phantom{.0}) \\ %
		\ButtonBench      & \text{total unique} & 49 & 14 & 11 (0.8) \\\midrule 
		\AsyncTaskBench   & \sharpsat-0.6 & 3 & 3 & 3 (1\phantom{.0}) \\ %
		\AsyncTaskBench   & \pfsa     & 13 & 10 & 4 (0.4) \\
		\AsyncTaskBench   & \hmm & 19 & 12 & 6 (0.5) \\ 
		\AsyncTaskBench   & \text{total unique} & 22 & 15 & 7 (0.5) \\\midrule
		all objects       & \text{total unique} & 171 & 75 & 55 (0.7) \\
        \bottomrule
\end{tabular*}

\end{table}

In \tblref{ManualTriage}, we show veracity results for the \sharpsat-0.6, \pfsa, and \hmm methods for each framework type.
We manually triaged the twenty rules with highest weight (using the parsimony assumption that the number of rules for each framework type should be small). We then add a row for the total number of unique rules in each column.  Since no single method dominated most of the time, this row shows that the combination of methods can learn a more complete specification than any single method alone.  In total we learned 75 unique rules which model the behavior of the framework. Of these, 55 directly correspond to the cause and effect that a developer would attribute to the system.

The highest level point is that a number of actual rules are learned for each framework type by both learning methods (the actual column) out of the thousands of possible rules shown in \tblref{objects} (7326) by only examining the top 20 rated rules from each learning method.
And furthermore, most actual rules that are learned correspond to the real root cause in the Android framework (the direct column). This observation provides supporting evidence for lifestate rules capturing event-driven object protocols at an appropriate level of abstraction.

The only exception to the high fraction of actual rules is the \AsyncTaskBench{} which can be fully specified by 7 rules as we describe further below. This biases the fraction to be low since we took the top 20 rated rules, and there were only 7 rules to learn. And every specification needed for \AsyncTaskBench{} was learned. Furthermore, most of the direct rules are found even in the top 10.

\newsavebox{\SBoxAsyncTaskonPostExecuteCi}\codejmk[\small]{\SBoxAsyncTaskonPostExecuteCi}{(t:AsyncTask).|ci:onPostExecute|()}

One remaining question is how many rules are required to specify the protocol.
Is it the case that the number of rules for each framework type should be small compared to the number of possible rules.
To address this question, we manually specified the full set of rules for \AsyncTaskBench{} for the observed events and callins (since this class is comparatively small with only 55 possible rules compared to the thousands possible for the other framework types, see \tblref{objects}).

Manual specification of \AsyncTaskBench{} resulted in 7 direct rules. The result is that we are learning 3 out of the 7 direct rules with the \sharpsat-0.6 method, 4 out of 7 with the \pfsa method, and 6 out of 7 with the \hmm method.  There were no rules in the manually created set which were not learned by one of our algorithms.

\paragraph{Threats to Validity.}

\JEDI{Traces.}
The rules that can be learned are limited by the observations available in the traces. For example, there was no opportunity to learn anything about the \codej{AsyncTask.cancel()} callin since our traces did not include any observations of such an invocation. In the end, we are limited by the extent by which framework behavior can be explored through dynamic execution. To try to mitigate this affect, we made traces with as many applications as possible minimizing duplicate traces.  This exposes us to more possible ways developers may use a given framework object.
%
\JEDI{Selecting framework types.}
There is a potential for bias in selecting the framework types on which to mine specifications.
To minimize this effect, we selected by looking just at the number and length of traces and diversity of events and callins (from \tblref{objects}).

\paragraph{Mined Specifications.}

From the veracity summary, the \sharpsat-0.6, \pfsa, and \hmm methods appear comparable, so we look more closely into the rules that are learned.
From our running example in \secref{overview}, the most important rule that we would expect is
that \codej{(t:AsyncTask).execute()} callin can only be called once.
It is indeed learned by all three methods.
Another important rule is that the \fmtevt{PostExecute} event is enabled by the \codej{(t:x AsyncTask).execute()} callin.
In this case, the \sharpsat-0.6 and \hmm methods learned this rule but the \pfsa method did not.


The biggest source of false rules learned that we observed are rules that appear to come from common patterns in how developers use the framework.
For example, one rule learned by \pfsa{} is that \codej{(b:Button).getWidth()} disallows itself, which perhaps corresponds to few traces with repeated calls to \codej{getWidth}.

A source of actual but not direct rules relates to callbacks and \codej{super} calls. We found that a common pattern in Android is to require a call to the super method.  For example to implement \codej{|cb:onActivityCreated|}, \codej{super.|ci:onActivityCreated|} would need to be called.
Thus in functioning applications, this \codej{super} call is seen as a callin that always happens with the event the triggers the \codej{|cb:onActivityCreated|} callback. Thus, the left-hand--side of a actual but not necessarily direct rule is where the left-hand--side is either the event or the \codej{super} callin. One could reasonably consider removing required \codej{super} calls from our definition of callins.

\newsavebox{\SBoxFragmentGetResource}\codejmk[\small]{\SBoxFragmentGetResource}{(f:Fragment).getResources()}

In the end, we found a rule that captures an undocumented behavior of Android in the official Android documentation.
The rule states that the \usebox{\SBoxFragmentGetResource} is allowed when the event \fmtevt{ActivityCreated} that triggers the \codej{|cb:onActivityCreated|} callback.
The \codej{getResources} callin will throw an exception if called before this \fmtevt{ActivityCreated} event.

\section{Related Work}
\label{sec:related-work}

\JEDI{Temporal specification mining}

\paragraph{Specification Mining.}
Our technique is broadly similar to mining typestate specifications for API interfaces from traces (e.g., \cite{whaley-spec-mining, acharya-fsm-mining, dallmeier-fsm-mining, gabel-fsm-mining, walkinshaw-ltl-fms-mining,popl-spec-mining,perracotta,DBLP:conf/kbse/PradelG09}). Our contribution can be seen as an extension that, in addition to inferring typestate properties, also infers the ordering constraints of lifecycle events that interleave and interact with typestate properties.
\citet{popl-spec-mining} presents one of the earliest attempts to infer temporal specifications of programs by mining observed traces. Notably, \citet{perracotta} extend the technique to scale to larger data-sets with imperfect traces, and \citet{DBLP:conf/kbse/PradelG09} improve the technique to mining specifications for \emph{collaborations} involving multiple, interacting objects.
There is also work on combining automata mining with value-based invariant mining~\cite{value-sequence-mining, efsm-models}.
However, we are unaware of any previous work that mines specifications for lifecycle events in addition to their interaction with typestate APIs for frameworks.


\JEDI{Rule based mining}
Other work also simplifies automata specifications by considering simpler rule-based approaches (e.g., \cite{dynamic-alternating-rules, static-alternating-rules}).
Subsequently, \citet{lo-qbec} extend this kind of specification to QBEC (quantified binary temporal rules with equality constraints) specifications.
However, these rule-based approaches differ from lifestate rules in that they express necessary causation (e.g for every \texttt{lock} there must be a subsequent \texttt{unlock}). In contrast, lifestate rules do not express such conditions, and these conditions are actually undesirable, as most events in event-driven systems \emph{enable} other events but do not \emph{cause} them to occur.


\JEDI{Dynamic spec mining}


\JEDI{Static spec mining}
Concerning static mining techniques,
\citet{alur-synthesis} propose a technique to infer a typestate interface for a given Java class. This technique could conceivably be used to statically infer typestate specifications for event-driven frameworks such as Android. However, their technique does not account for the unique relationship between typestate and lifecycle events in event-based systems.
Additionally, \citet{shoham-yehav} develop a technique for inferring the typestate of a certain API by analyzing presumably correct client programs. The order of method invocations, determined statically, can produce a multi-object typestate specification. \citet{ramanathan-static-specs} propose a similar technique, but one that can also derive invariant-based preconditions. However, to our knowledge these techniques also cannot be successfully extended to event-based systems, where typestate and lifecycle properties may interact.


\paragraph{Analysis of Event-Driven Systems.}


Recently, more and more program analyses have focused on event-driven systems. Both dynamic analyses, such as for race detection~\cite{droidracer, cafa}, and static analyses for information flow~\cite{DBLP:conf/pldi/ArztRFBBKTOM14} and safety properties~\cite{DBLP:conf/oopsla/BlackshearCS15} expect some specification of framework behavior with respect to callbacks.
There has also been a number of recent static analyses targeting information flow concerns (e.g., 
\cite{DBLP:conf/ndss/GordonKPGNR,DBLP:conf/sigsoft/FengADA14,DBLP:conf/ccs/WeiROR14}) where a significant concern is synthesizing models for taint flow through the framework~\cite{DBLP:conf/popl/BastaniAA15}.



There are also some tools concerned with exposing the implicit control flow from the framework to callbacks by linking
callback registration methods with their callbacks~\cite{DBLP:conf/ndss/CaoFBEKVC15,node-js-cg},
modeling GUI components~\cite{DBLP:conf/icse/YangYWWR15}, and explicating reflection~\cite{DBLP:conf/pldi/BlackshearGC15}.
This information is useful for creating an over-approximation of the possible callbacks but does not capture precise events that may enable or disable a given event that then invokes the callback.



\section{Conclusion}
\label{sec:conclusion}

We have presented \emph{lifestate rules}, a language for specifying event-driven object protocols.
The key idea underlying lifestate rules is a model of event-driven programs in terms of enabled events and allowed callins unified as suspended messages.
As a result, lifestate rules are able to capture mixed lifecycle and typestate constraints at a higher level abstraction than automata.
We then instantiated this model in a dynamic analysis tool called \toolname{} that prepares Android traces for specification mining
using a notion of trace slicing adapted to events and callbacks.
Finally, we applied specification mining techniques to infer lifestate specifications based on unsupervised automata-based learning techniques that have been previously applied to typestate mining, as well as a direct lifestate mining technique based on propositional model counting.
In the end, we were able to learn several actual lifestate rules that
accurately capture the behavior of Android in a compact way.




\bibliographystyle{abbrvnat}

\bibliography{conference.short,bec.short,main.short}

%
%

\end{document}

%% file: abstract.tex
We present \emph{lifestate rules}---an approach for abstracting event-driven object protocols.
Developing applications against event-driven software frameworks is notoriously difficult.
One reason why is that to create functioning applications, developers must know about and understand the complex protocols that abstract the internal behavior of the framework.
Such protocols intertwine the proper registering of callbacks to receive control from the framework with appropriate application programming interface (API) calls to delegate back to it.
Lifestate rules unify lifecycle and typestate constraints in one common specification language. Our primary contribution is a model of event-driven systems $\semname$ from which lifestate rules can be derived.
We then apply specification mining techniques to learn lifestate specifications for Android framework types.
In the end, our implementation is able to find several rules that characterize actual behavior of the Android framework.

%% file: sharpsat.tex
%
%
%
In contrast to the two-phase probabilistic automata approaches,
we propose an algorithm that directly learns a lifestate specification from the
set of signature traces $\abstraceset$ that is potentially easier to
interpret.

The learning algorithm proceeds by computing a weight for each
possible rule in $\RuleSet$ using the traces in $\abstraceset$ and
then obtains a lifestate specification by selecting
the rules that meet a weight threshold.
The weight assigned to a rule is an average, among all the traces in
$\abstraceset$, of the fraction of the total executions that are
sound for that rule.

Intuitively, the ideal settings to compute the weight is to
observe the sequence of signature states and transitions
(a \emph{signature path}).
The main difficulty we face is that the internal state of the
permitted messages (i.e., the signature state
$\enSigState[\absof{\trace}][\absof{\eventmap}][\absof{\callinmap}]$)
is not observable.
To address the problem, we propose an abstraction of
the signature paths, which we use in our learning algorithm.

We assume no prior knowledge on the internal behavior of the
framework and define the most conservative abstraction via the \emph{non-deterministic 
transition relation} $\jstep[\nondet]{ \absof{\state} }{ \absof{\state}' }$ that, upon
invoking a \emph{permitted} enable or allow signature transition $\absof{\trans}$, can
\emph{non-deterministically} change the internal state of the system 
(i.e., permits or prohibits any other message) 
but keeps the semantics of disallow and disables as in Figure~\ref{fig:specification}.  
This choice of abstraction may produce a large number
of spurious behaviors that could affect
the quality of rules we learn, but we make the following observations:
\begin{inparaenum}[(i)]%
\item the imprecision introduced by the spurious transition
may be reduced by observing a large number of traces;
\item with a further knowledge of the framework, we can
reduce the non-determinism (i.e. reducing the abstract paths),
hence improving the quality of the learned specification.
\end{inparaenum}
%
%

The main challenge with this approach is that enumerating all abstract 
paths for a signature trace is intractable. To overcome this problem, 
we represent the set of possible abstract paths symbolically
encoded as a propositional logic formula, to leverage 
efficient model counting solvers.

\subparagraph{Learn Specifications by Counting Paths.}
%
Given a signature trace
$\abstrace\colon \absof{\trans}_1 \ldots \absof{\trans}_n$, let
$\AbsPaths{\abstrace}$ denotes the set of all the sequences of
states and transitions (\emph{paths}) under the non-deterministic
semantics.
%
For each rule kind $\specEnable{\absforce}{\absmsg}$
(resp. $\nspecArrow \specDisableOp, \specArrow \specAllowOp, \nspecArrow \specDisallowOp$),
we say that the rule is ``matched'' in a state of the path if, in that state,
the message $\absmsg$ is enabled (resp. disabled, allowed, disallowed)
and the abstract transition executes the message $\absforce$.

We want to count all the paths of a trace such that the ratio of the number of
states of the path that match the rule over the total number of times we seen
$\absforce$ in the trace is greater than a weight $w$.
For a signature trace $\abstrace$, we
call \(\AbsPathsRule{\abstrace}{\absrule}\) such set of paths.

%
%
Given a signature trace $\abstrace$ and a rule $\absrule$,
the \emph{frequency} of $\absrule$ in $\abstrace$ with weight
$\weight$, is given by:
\[
\TraceFreq{\weight}{\absrule}{\abstrace} \defeq
\frac{\Card{\AbsPathsRule{\abstrace}{\absrule}}}{\Card{\AbsPaths{\abstrace}}}\;.
\]
%
%
The notion can naturally be lifted to a set of signature traces
$\abstraceset$ by taking the geometric mean of the fractions of the
observed traces.

Let $\delta$ be a real constant. The final specification is
\(
\absof{\directiveset} \defeq \{\absrule\ |\ \TraceFreq{\weight}{\absrule}{\abstraceset} \ge \threshold \}
\).
The threshold $\delta$ is needed to select only the rules that, according to
the frequency value, more likely capture the framework behavior.


The number of paths in the sets $\AbsPaths{\abstrace}$ and
$\AbsPathsRule{\abstrace}{\absrule}$ is exponential in the total
number of messages.
Thus, it is not feasible to explicitly enumerate them.
Our solution is to encode all the paths in
$\AbsPaths{\abstrace}$ as a propositional logic formula.
The intuition is that each (complete) model of this formula
represents a path in $\AbsPaths{\abstrace}$.
Hence, we cast the problem of counting the cardinality of
$\AbsPaths{\abstrace}$ to the problem of counting the number of
models of the Boolean formula, for which several
efficient tools exist (e.g.~\cite{sharpsattool}).
The encoding is an adaptation of Bounded Model Checking~\cite{bmc},
a technique that encodes all the paths of length $k$ of a transition
system.